\documentstyle[12pt,epsf,wrapft,mbf]{article}

\newcommand{\BE}{\begin{equation}}
\newcommand{\EE}{\end{equation}}
\newcommand{\BA}{\begin{eqnarray}}
\newcommand{\EA}{\end{eqnarray}}

\newcommand{\vk}{\mbf k}
\begin{document}
\begin{flushright}TWC-97-1\end{flushright}
\vspace{1cm}

\begin{center}
\begin{Large}
{\bf Transverse Photon Spectrum from QGP Fluid}
\end{Large}

\vspace{1.cm}

\begin{large}
Tetsufumi  Hirano, Shin Muroya$^*$,and Mikio Namiki
\end{large}
\vspace{1.cm}

{\it Department of Physics, Waseda University, Tokyo 169 }\\
$^*${\it Tokuyama Women's College, Tokuyama, Yamaguchi 754 \\}
\end{center}

\vspace{.5cm}

\begin{abstract}
We calculate the thermal photon
distribution from the hot QCD
matter produced by high energy nuclear collisions based on a 
hydrodynamical model,
 and compare it with the recent experimental
data obtained by CERN WA80.
Through the asymptotic value of the slope
parameter of the transverse momentum distribution, we 
investigate the characteristic temperature of the QCD fluid.
\end{abstract}



\vspace{.5cm}

\noindent
\begin{large}
{\bf Introduction}
\end{large}
\vspace{.5cm}

Since the thermal photon is considered to keep the information
about the early stage of the hot matter produced by relativistic
nuclear collisions, many theoretical analyses have already been
done.  Many groups \cite{Srivastava} have
analyzed the experimental data of CERN WA80 S+Au
200GeV/nucleon \cite{WA80p} so as 
to fit their theoretical model to the thermal photon
emission data.  

In this paper  we analyze the photon and the hadron
distribution produced by the hot QCD matter
in a consistent way \cite{Hirano}:
\begin{enumerate}
\item We first choose parameters in the hydrodynamical
model so as to reproduce the hadronic spectrum, {\it i.e.},
the (pseudo-)rapidity distribution and the transverse momentum
distribution.
\item We derive the thermal production rate of photons from a
unit space-time volume based on the finite temperature 
field theory.
\item Accumulating the thermal production rate, 
 over the whole
space-time region covered with the particle source, which is 
estimated by the hydrodynamical model,
we evaluate the thermal photon distribution which is to
be compared with the experimental data.
\end{enumerate}


\vspace{.5cm}
\noindent
\begin{large}
{\bf Hydrodynamical Model and Hadronic Spectrum}
\end{large}
\vspace{.5cm}

In a previous paper \cite{Muroya95}, 
  by making use of the following two models:
I) the QGP fluid model with phase transition,
I$\!$I) the hot hadron gas model without phase transition,
we \\
have analyzed the pseudo-rapidity distribution
of charged hadrons in S+Au 200 GeV/nucleon collision
obtained by CERN WA80 \cite{WA80h}. 
Where we supposed that the fluid in the QGP phase is
dominantly composed of u-, d-,
s-quarks and gluons and that the fluid in the hadron phase
is dominantly composed  
of pions and kaons. 

According to the previous analysis \cite{Muroya95},
we here use the first model (the QGP fluid model with phase
transition) specified by the
initial temperature
$T_i=195$~MeV, the critical temperature $T_c=160$~MeV,
and the freeze-out temperature $T_f=140$~MeV, and the
second model (the hot hadron gas model without phase transition)
specified by $T_i=400$~MeV and $T_f=140$~MeV.
For these models, we obtain theoretical
results of the hadronic spectrum.
From Fig.~1 and Fig.~2, 
we observe that the both models can consistently reproduce the
experimental data \cite{Muroya95}. 

\noindent
\begin{minipage}[t]{8cm}
\begin{wrapfigure}[20]{l}{7.6cm}
\epsfxsize=6cm
\centerline{\epsfbox{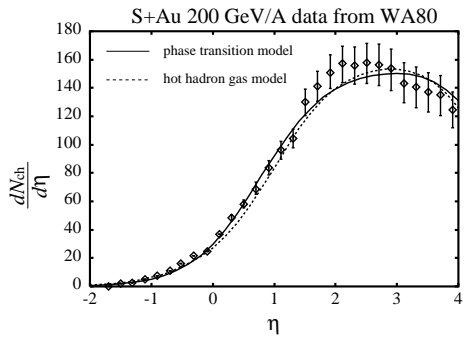}}
\caption{ The pseudo-rapidity distribution of charged
hadrons in S+Au 200 GeV/nucleon collision.  Data form
 CERN WA80. The solid curve and the dashed
curve  stand for, respectively, the QGP phase transition model
 and the hot hadron gas model.}
\label{fig:2}
\end{wrapfigure}
\end{minipage}
\begin{minipage}[t]{8cm}
\begin{wrapfigure}{l}{7.6cm}
\epsfxsize=6cm
\centerline{\epsfbox{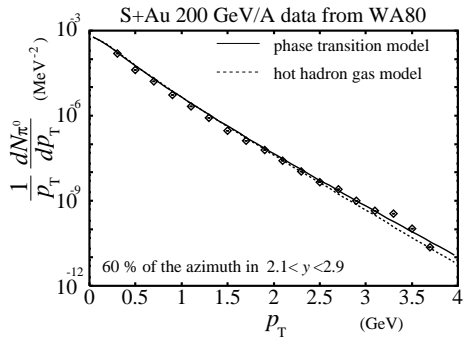}}
\caption{The transverse momentum distribution of neutral
pions in S+Au 200 GeV/nucleon collision. Data was
obtained by CERN WA80.  The solid curve
and the dashed
curve stand for, respectively, the QGP phase transition
model and the hot hadron gas model.
}
\label{fig:3}
\end{wrapfigure}
\end{minipage}\\

\vspace{8.5cm}

\noindent
\begin{large}
{\bf Thermal Production Rate of Photons}
\end{large}
\vspace{.5cm}

Assuming that a certain mode is dominantly excited in a
local equilibrium system of the hot QCD matter and
that the canonical operator of that mode obeys the
{\it quantum} Langevin equation \cite{Mizutani88}, we can
easily derive thermal production rate semi-phenomenologicaly.
In Fig.~3 we compare the numerical result of our
semi-phenomenological
production rate
with another result obtained by Kapusta {\it et
al.} \cite{Kapusta,Hirano}. 

Integrating the production rate from a volume element $R(T)$ 
over the whole space-time volume 
in which the particle source exists, we obtain momentum
distributions
\BE \label{inv}
\left.
k_0 \frac{d^3 N}{d {\vk}^3} = \int d^4 x
k_0 ^\prime\frac{d^3 R(T(x))}{d {\vk}^{\prime 3}}
\right| _{k_0 ^\prime = U^\mu(x)k_\mu},
\EE
which are to be compared with experimental data.
Here temperature $T(x)$ and local four velocity $U^\mu(x)$ 
at space-time point $x$ are
given by the numerical solution of the hydrodynamical model.  
Figure 4 shows the numerical results of Eq.~({\ref{inv}) 
compared with the recent 
experimental data (S+Au 200 GeV/nucleon collision)
obtained by CERN WA80 \cite{WA80p}. 
The solid curve and the dotted curve are, respectively, 
the whole thermal photon distribution given by our QGP fluid
model and the contribution of the QGP phase region only.
In the case of the QGP fluid with
phase transition model, our result in Fig.~4 seems consistent with
the experimental data of WA80. 
The dashed curve stands for the photon distribution given by 
our hot hadron gas model.
The dashed curve deviates from the
experimental data in both absolute value and slope. 
\\
\begin{minipage}[t]{8cm}
\begin{wrapfigure}{l}{7.9cm}
\epsfxsize= 7.5 cm
\centerline{\epsfbox{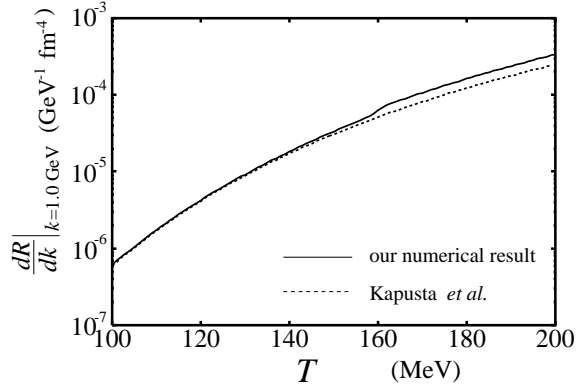}}
\caption{The production rate as a function of temperature
. The solid curve
stands for our phase transition model
and the dashed curve for the production rate 
calculated in Ref.[4].
The critical temperature $T_c=160$ MeV.}
\label{fig:5}
\end{wrapfigure}
\end{minipage}
\begin{minipage}[t]{8cm}
\begin{wrapfigure}{l}{7.9cm}
\epsfxsize= 7.5cm
\epsfysize= 7.5cm
\centerline{\epsfbox{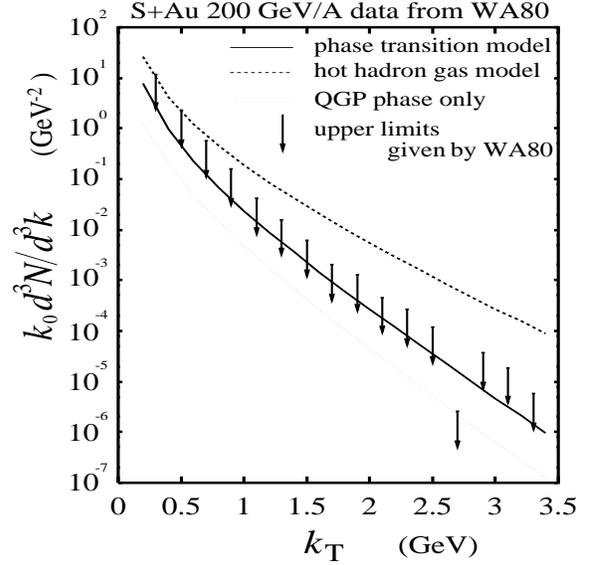}}
\caption{The single photon spectrum
in S+Au 200 GeV/nucleon collision.
}
\label{fig:6}
\end{wrapfigure}
\end{minipage}\\
\vspace{9.0cm}

\noindent


\noindent
\begin{large}
{\bf Effective temperature }
\end{large}
\vspace{.5cm}

In order to pick up the most dominant contribution to the
transverse momentum distribution,
we can rewrite the thermal factor as
\BE
\left. \exp(-\frac{k_\mu
U^\mu}{T})\right| _{\mbox{
\begin{small}
c.m. system
\end{small}
}}
\Longrightarrow 
\exp(-\frac{k_T \frac{1-v_T}{\sqrt{1-v^2}}}{T})
= \exp(-\frac{k_T}{\sqrt{1-\frac{v_L^2}{1-v_T^2}}
\sqrt{\frac{1+v_T}{1-v_T}}T}),
\EE
by which we can define the effective temperature $T_{eff} $
of the fluid at the 
volume element with velocity $v_L, v_T$ and temperature $T$
as,
\BE
 \label{teff}
T_{eff} = 
\sqrt{1-\frac{v_L^2}{1-v_T^2}}\sqrt{\frac{1+v_T}{1-v_T}}T .
\EE

Table~I shows the maximum $T_{eff}$
given by our numerical results of hydrodynamical model simulation
and the slope parameter at $k_T = 20$ GeV for the above two models.
Through comparison of $T_{eff}$
with the slope parameter $T_s$ in Table~I, we know that
the asymptotic slope parameter $T_s$ has possible
origins different from each other for the above two models:
The critical temperature dominates in the QGP fluid model 
with phase transition,
while the initial
temperature dominates in the hot hadron gas model without phase transition.

\begin{small}
\begin{table}
\caption{The maximum $T_{eff}$ in our hydrodynamical simulation, and the slope
parameters $T_s$ at $k_T$ = 20 GeV.
}
\label{table:1}
\begin{center}
\begin{tabular}{c|c|c|c|c|c}
\hline \hline
Model & $T$ (MeV)& $v_T$ & $v_L$  & $T_{eff}$  (MeV) & $T_s$ (MeV)\\
\hline
QGP fluid (QGP phase) & 157.5 & 0.53 & 0.11 & 280.8 &
273.8\\
QGP fluid (hadron phase) & 157.5 & 0.53 & 0.11 & 280.8 & 
273.5\\
\hline
Hot hadron gas & 400.0 & 0.0 & 0.0 & 400.0 & 397.0\\
\hline
\end{tabular}
\end{center}
\end{table}
\end{small}

\vspace{.5cm}
\noindent
\begin{large}
{\bf Concluding Remarks}
\end{large}
\vspace{.5cm}

We have derived the thermal photon distribution emitted
from a hot matter produced by the high energy nuclear collisions, 
based on hydrodynamical model.  
We have observed that only the QGP fluid (with phase
transition) model can consistently reproduce
 S+Au 200 GeV/nucleon data obtained by CERN WA80
experimental data.
Furthermore we have discussed the asymptotic slope 
parameter of the transverse photon distribution.

The authors are much indebted to Professor I.~Ohba, H.~Nakamura, and 
other members of 
high energy physics group of Waseda Univ. for their helpful 
discussion. 

\begin{small}

\end{small}
\end{document}